\documentclass[lettersize,journal]{IEEEtran}
\usepackage{amsmath,amssymb,amsfonts}
\usepackage{amsthm}
\usepackage{algorithmic}
\usepackage{algorithm}
\usepackage{array}
\usepackage[caption=false,font=normalsize,labelfont=sf,textfont=sf]{subfig}
\usepackage{textcomp}
\usepackage{stfloats}
\usepackage{url}
\usepackage{verbatim}
\usepackage{graphicx}
\usepackage{cite}
\usepackage{arydshln}
\usepackage{multirow}
\usepackage{xcolor}
\usepackage{threeparttable}

\newtheorem{proposition}{Proposition}

\newcommand{\tabincell}[2]{\begin{tabular}{@{}#1@{}}#2\end{tabular}}
\begin{document}

\title{Optimal Pricing of Electric Vehicle Charging on Coupled Power-Transportation Network based on Generalized Sensitivity Analysis}

\author{
Lyuzhu~Pan,~\IEEEmembership{Graduate Student Member,~IEEE,}
Hongcai~Zhang,~\IEEEmembership{Senior Member,~IEEE,}
Yan~Xu,~\IEEEmembership{Senior Member,~IEEE}
\thanks{
This paper is funded in part by the National Natural Science Foundation of China (File no. 52007200) and in part by the Science and Technology Development Fund, Macau SAR (File no. 0094/2024/AGJ, and File no. 001/2024/SKL) (Corresponding author: \textit{Hongcai Zhang}.)

L. Pan, and H. Zhang are with the State Key Laboratory of Internet of Things for Smart City and Department of Electrical and Computer Engineering, University of Macau, Macao, 999078 China (email: hczhang@um.edu.mo).

Y. Xu is with the Center for Power Engineering, Nanyang Technological University, Singapore 639798. (e-mail: xuyan@ntu.edu.sg).}
}



\maketitle

\begin{abstract}
In the last decade, charging service providers are emerging along with the prevalence of electric vehicles. These providers need to strategically optimize their charging prices to improve the profits considering operation conditions of the coupled power-transportation network. However, the optimal pricing problem generally involves the user equilibrium model, which leads to a mathematical program with equilibrium constraints. As a result, the pricing problem is non-convex and computationally intractable especially for large-scale network. To address this challenge, we propose a generalized sensitivity analysis approach for optimal pricing of electric vehicle charging on coupled power-transportation network. Specifically, we adopt a sensitivity analysis to capture the best response of charging demand to charging price in the gradient form. Consequently, charging service providers can make pricing decisions based on the gradient information instead of the conventional KKT conditions of the user equilibrium model. We then propose a tailored gradient descent algorithm to solve the whole pricing problem. The mathematical proof of validity is given and the time complexity of the proposed algorithm is theoretically polynomial. Numerical experiments on different scales of networks verify the computational efficiency of the proposed algorithm, indicating its potential in evaluating the impact of the optimal pricing on the operational performance of large-scale coupled power-transportation network.

\end{abstract}


\section{Introduction}
\IEEEPARstart{T}{he} greenhouse gas emissions from the transportation sector significantly contribute to climate change. For instance, in the United States, the transport sector accounts for 29\% of total greenhouse gas emissions\cite{transportation}. Therefore, to mitigate global warming and achieve a net-zero society, transportation electrification has become a key decarbonization pathway for countries and regions such as China\cite{ChinaCarbonPeak}, the United States\cite{USCarbonBlueprint}, the European Union\cite{EU2023transport}, etc. By the end of 2023, the global stock of electric vehicles (EVs) had nearly increased eightfold compared to 2018, reaching 40 million\cite {globalEVtrend}. Moreover, EVs require more frequent refueling than gasoline vehicles due to their limited battery capacity, leading to a soaring deployment of fast charging stations (FCSs). The increasing FCSs and charging power have been coupling both power and transportation networks together. As a result, unregulated large-scale charging behaviors may lead to the negative impact on the coupled power and transportation network\cite{zhang2024sustainable}.

{In recent years, extensive efforts have been devoted to the operation of the coupled network through pricing strategies. These studies can be summarized as two types in terms of different stakeholders: the \textit{social operator} and the \textit{charging service providers}.
From the perspective of the social operator, some tariff-based policies can guide EVs' behaviors for better social welfare. For instance, locational marginal prices of the power network are widely adopted as charging prices\cite{zhang2020power,lv2021integrated,shao2023generalized}, as they can guide rational EV drivers to recharge at the appropriate power buses. Therefore, the loss and congestion of the power distribution network are improved. The above studies focus on the performance of the power network while neglecting the welfare of the transportation network. For the transportation network, there is a motivation to reduce overall time costs through various approaches, with traffic tolls being the most common\cite{sumalee2011Marginal}. Some works are conducted to coordinate the social welfare of the coupled network. For example, Sheng \emph{et al. }\cite{sheng2021coordinated} and Xie \emph{et al. }\cite{xie2023collaborative} adopt locational marginal prices as charging fees, while also imposing road tolls to guide traffic flows. To alleviate the congestion of FCSs, Li \emph{et al. }\cite{li2023inverse} further impose a plug-in fee to EV drivers when they recharge EVs at crowded FCSs. }

{In practice, most public charging stations are owned by self-interested charging service providers, meaning that the social operator cannot stipulate charging prices or impose plug-in fees. Consequently, a growing number of studies concentrate on optimal or competitive pricing strategies from the perspective of charging service providers.} {For instance, Chen \emph{et al. }\cite{chen2022trilevel} maximize the profit of the charging station with an approximated charging demand curve. Lai \emph{et al. }\cite{lai2024bargaining} propose a Nash–Harsanyi game framework to optimize the charging price strategies in the cooperative environment. Although the above studies for charging service providers describe how drivers are sensitive to distance and charging prices, they cannot fully consider the transportation topology. To simultaneously account for how the topology of the transportation network affects drivers' decisions and how the EV drivers respond to the charging price variation, some researchers incorporate the user equilibrium model in the pricing problem \cite{yang2025recent}. The user equilibrium model can depict the EV drivers' routing and charging behaviors in the transportation topology.} {For instance, Yan \emph{et al. }\cite{yan2024hierarchical} propose a hierarchical game model to optimize the charging prices, where the transportation model is the user equilibrium. However, when the user equilibrium model is incorporated into the pricing problem, it also introduces intractable bilinear terms and nonlinear complementary constraints (see details in Section \ref{complexity analysis}). As a result, the pricing problem becomes computationally intractable, especially for large-scale problems. To tackle computational complexities, current studies employ two types of methods: deterministic method and heuristic method. The deterministic method simplifies the original problem by using various relaxation techniques. For instance, Li \emph{et al. }\cite{li2023strategic} linearize the bilinear term in the objective function via the McCormick envelope at the cost of optimality. Sheng \emph{et al. }\cite{sheng2023impact} discretize the charging price and reformulate the bilinear term as the combination of binary variables, which provides a more accurate approximation but worse efficiency.} {Lv \emph{et al. }\cite{lv2024optimal} design a relaxation-based iterative algorithm to handle the large-scale complementary constraints, instead of the big-M method using integers. Zeng \emph{et al. }\cite{zeng2023conic} address the bilinear term using the strong duality property of the user equilibrium problem. However, the relaxation of complementary constraints relies on the stochastic user equilibrium model where path flows are forced to be positive. The heuristic method is based on meta-heuristic or machine learning algorithms. Cui \emph{et al. } \cite{cui2021optimal,cui2023multiperiod} combine the genetic algorithm and mathematical programming to solve the competitive pricing problem in static and dynamic scenarios, respectively. Qian \emph{et al. }\cite{qian2022multiagent}, Ye \emph{et al. }\cite{ye2023identifying} and Yang \emph{et al. }\cite{yang2024multiagent} apply deep reinforcement learning to solve the charging service pricing problem. However, these methods are widely criticized for issues related to convergence, optimality, and computational efficiency.}

In this study, we propose an optimal pricing model from the perspective of charging service providers, where the topology of the transportation network is fully considered and the response of charging demand to charging price is captured via the user equilibrium model. We then adopt the clearing electricity price from the power network as the cost of charging service providers. To cope with the computational complexities, we propose a comprehensive solution framework based on the gradient descent algorithm, wherein the drivers' response is captured in the form of the gradient of the user equilibrium via the generalized sensitivity analysis. {To clarify our contributions, we summarize and compare this study with other literature in terms of pricing scheme, system model, and solving methods, which is presented in Table \ref{tab:literature}.} The contributions of our paper are twofold:
\begin{table}[]
\centering
    \caption{{Comparison with Related Literature}}
    \label{tab:literature}
\begin{tabular}{ccccc}
\hline
Ref. &  \tabincell{c}{Power\\network} & \tabincell{c}{Transportation\\network} & Method \\ \hline
\cite{chen2022trilevel} & \checkmark& & \tabincell{c}{Constraint generation algorithm\\\& discretization relaxation}  \\ \hline
\cite{lai2024bargaining}   & \checkmark& &Interior point optimizer \\ \hline
\cite{yan2024hierarchical,li2023strategic} & \checkmark & \checkmark &       \tabincell{c}{Mixed-integer programming\\\& McCormick relaxation}\\ \hline
 \cite{sheng2023impact} &   & \checkmark  &  \tabincell{c}{Mixed-integer programming\\\& discretization relaxation}     \\  \hline
  \cite{zeng2023conic} &   & \checkmark  &  \tabincell{c}{Penalty convex-concave\\procedure \& conic relaxation}     \\ \hline
\cite{cui2021optimal,cui2023multiperiod} & \checkmark  & \checkmark  &  \tabincell{c}{Combined genetic algorithm\\\& mathematical programming}  \\ \hline
\cite{ye2023identifying,yang2024multiagent} & \checkmark  & \checkmark  &  Deep reinforcement learning  \\ \hline
 \tabincell{c}{This\\work}& \checkmark  & \checkmark  &  \tabincell{c}{Gradient descent algorithm\\\& generalized sensitivity analysis}  \\
 \hline
\end{tabular}
\end{table}
\begin{enumerate}
    \item We introduce the sensitivity analysis technique into the optimal pricing problem to effectively capture the rational response of EV drivers to charging prices using gradient information. This approach needs not to embed the KKT conditions of the user equilibrium model into the pricing problem, thus avoiding the complex non-convex formulation. Owing to the gradient information, the pricing problem can be addressed by iteratively solving two convex problems based on the proposed tailored gradient descent algorithm. As a result, the proposed algorithm shows significant computational superiority over conventional mathematical programming.
    \item Considering the limitation of the conventional sensitivity analysis in terms of accommodating the charging behaviors of EVs, we generalize it via the hyper-arc transformation, which preserves the same mathematical property of the original problem. We prove that the Jacobi matrix of the KKT system of the user equilibrium model is still non-singular under the proposed generalization, which further guarantees the validity of the generalized sensitivity analysis. {We also prove that the time complexity of the proposed gradient descent algorithm is polynomial with the number of path set and Origin-Destination pairs.}
\end{enumerate}
\noindent In addition, numerical experiments are carried out to validate the algorithmic performance of the proposed method. The result validates the effectiveness of the proposed algorithm. 

The rest of the paper is organized as follows. We state the problems and analyze the complexities of the common mathematical programming method in Section \ref{problem statement}. We then introduce the proposed solution method in detail in Section \ref{solution method}. Numerical experiments are carried out in Section \ref{result}. The comprehensive conclusions are made in Section \ref{conclusion}.

\section{Problem Statement}\label{problem statement}
\begin{figure}
\centerline{\includegraphics[scale=0.25]{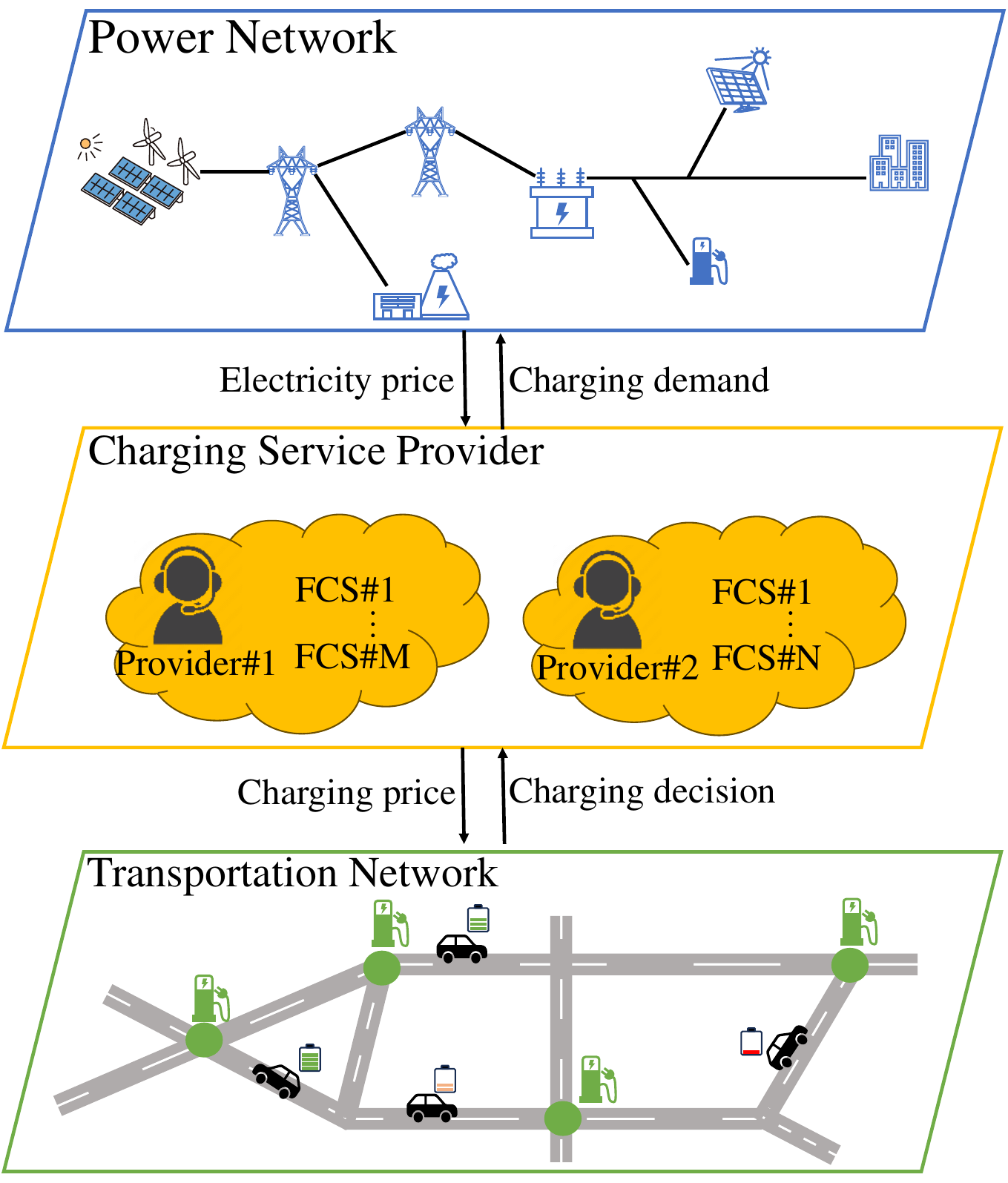}}
\caption{Structure of the optimal pricing problem.}
\label{fig:CPTN}
\end{figure}
The relationship between different entities in this study is demonstrated in Fig.~\ref{fig:CPTN}. From the perspective of the charging service provider, it stipulates charging prices based on three factors: EV drivers' charging decisions, electricity prices from the power network\footnote{It should be noted that we do not introduce the power network dispatch model in the subsequent formulations since our main contribution focuses on adopting a new charging demand capture method. {The detailed interaction between the power network and the charging service provider is presented in Appendix \ref{appendix:power network}.}}, and rivals' prices\footnote{{Note that the price competition between different providers is not explicitly modeled since the focus of this study is designing efficient algorithm for the optimal pricing problem. Instead, we optimize the price strategy of the given provider, while rivals' prices are assumed to be constant. Nevertheless, our model can be easily extended to a non-cooperative game model and solved with the Gauss-Seidel iteration algorithm (see details in reference \cite{sheng2023impact}).}}. {The interactions in our model can be summarized as follows. Firstly, rational EV drivers in the transportation network make the optimal routing and charging decisions according to charging prices. Secondly, the power network adjusts the locational marginal price as the electricity price in response to the charging demand. Thirdly, the charging service provider should strategically stipulate prices considering the pricing strategies of other providers.} 

{The proposed model can be adopted for the urban-scale optimal pricing, involving both power and transportation networks, price competition between charging service providers, and large EV populations. On the one hand, this model can provide an optimal pricing strategy to greatly improve the profits of charging service providers. On the other hand, policymakers can adopt this model to evaluate the impact of the charging price competition on the urban power and transportation flow. In practice, the power and transportation networks are extremely complex, and the scale of public charging stations is large, such as New York \cite{chargefinder}, resulting in the intractable optimal pricing problem. To address the computational burden of large-scale scenarios in the real world, we propose an efficient sensitivity analysis based algorithm. }

{It should be noted that we adopt the static user equilibrium model in this study for more concise analytical derivation. Nevertheless, the proposed model can be transformed into semi-dynamic form via repeatedly invoking it over consecutive time intervals (see details in reference \cite{lv2019optimal}). Since the EV behaviors are modeled based on aggregated traffic flows, the EV drivers' privacy can be ignored and their behavior uncertainties in the large-scale problem is negligible. As a result, we do not account for the uncertainties in the model, which is common in the related studies \cite{cui2021optimal,li2023strategic,ye2023identifying}. Besides, it is assumed that EVs need to recharge only once during the trip with the same charging energy $E$ in kWh in this model. Although it seems to be a strong assumption, it is acceptable in a macro-analysis study. Additionally, the rationality of this assumption has been well justified in the reference \cite{wei2018network} and widely adopted in related papers, such as \cite{sheng2021coordinated} and \cite{ye2023identifying}.  Furthermore, all formulations in the main body only consider EVs to facilitate the understanding of the algorithm derivation. For generalization, we attach the formula derivation with both EVs and the conventional internal combustion engine vehicles in Appendix \ref{appendix:mixed}.}

In this section, we first construct the user equilibrium model and the charging service pricing model. We then reformulate these two models into a single-level model and analyze its complexities.  
\subsection{User Equilibrium Model}\label{uemodel}
The user equilibrium model is a classical tool in transportation studies that depicts how rational drivers seek the ``shortest" path in terms of generalized travel costs (e.g., time and fuel costs). The results of the user equilibrium model indicate a state where drivers have found their shortest paths, and no driver can reduce their travel cost by unilaterally changing their routes\cite{sheffi1984urban}.

\begin{figure}[]
\centerline{\includegraphics[scale=0.5]{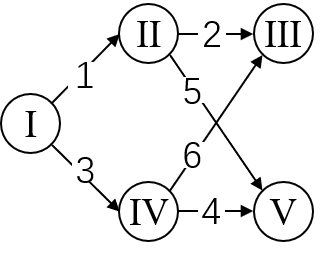}}
\caption{Illustrative transportation network network. The Roman number in the circle represents the index of the node. The number on the arrow is the arc index. Assume $\mathcal{N}^\text{fcs}=\{II,III\}$, $\mathcal{W}=\{(I,III,1.5),(I,V,2.0)\}$.}
\label{illustration}
\end{figure}

The transportation network is represented by a directed graph $G_{\text{TN}}=(\mathcal{N}, \mathcal{A})$ consisting of a set of nodes $n\in\mathcal{N}$ and arcs $a\in\mathcal{A}$. EVs can recharge at a subset of the transportation nodes $m\in\mathcal{N}^{\text{fcs}}\subset \mathcal{N}$, which are assumed to be equipped with FCSs. The arc is the road segment connecting two transportation nodes. To depict the traffic demand, we use the concept of the Origin-Destination (OD) pair. The OD pair is denoted by a tuple $(s,t,d)_{w}\in\mathcal{W}$, where $d$ is the traffic demand from origin $s\in\mathcal{N}$ to destination $t\in\mathcal{N}$, $\boldsymbol{D}=[d_w]\in\mathbb{R}^{|\mathcal{W}|}$ is the traffic demand vector and $\mathcal{W}$ denotes the OD pair set. For every OD pair, there are several paths guiding drivers and they are stored in the arc-path matrix $\Delta^\text{arc}$. Each column of $\Delta^\text{arc}=[\delta_p]$ ($\delta\in\mathbb{M}^{|\mathcal{A}|}$) is called a path consisting of arcs that it will pass through. We let vector $\mathcal{P}=[p]$ as the path set storing the column index of $\Delta^\text{arc}$. In order to distinguish the paths of different OD pairs, we introduce an auxiliary matrix $\Lambda\in\mathbb{M}^{|\mathcal{W}|\times|\mathcal{P}|}$ called OD-path matrix. We then use a charge-path matrix $\Delta^\text{fcs}\in\mathbb{M}^{|\mathcal{N}^{\text{fcs}}|\times|\mathcal{W}|}$ to indicate the charging selection of each EV path. Note that bold symbols denote the vectors or matrices, $|\cdot|$ denotes the cardinality of the set, $\mathbb{R}$ and $\mathbb{M}$ are the real number and binary set, respectively. Here we give an illustrative example (see Fig.~\ref{illustration}) to help understand the concept of matrices.

According to the illustrative case, we write the matrices below. For path (column) \#1, it passes arcs\#1 and \#2 (see the first matrix $\Delta^\text{arc}$) and recharges at FCS\#1 located at node II (see the second matrix $\Delta^\text{fcs}$). In the third matrix $\Lambda$, there are four paths, wherein the first two paths are associated with the first OD pair and the last two paths are associated with the second OD pair.
\begin{align*}
    \Delta^{\text{arc}} &= 
    \begin{bmatrix}
        1 & 0 & 1 & 0 \\
        1 & 0 & 0 & 0 \\
        0 & 1 & 0 & 1 \\
        0 & 0 & 0 & 1 \\
        0 & 0 & 1 & 0 \\
        0 & 1 & 0 & 0 \\
    \end{bmatrix},\\
    \Delta^{\text{fcs}} &= 
    \begin{bmatrix}
        1 & 0 & 1 & 0 \\
        0 & 1 & 0 & 1 \\
    \end{bmatrix},\\
     \Lambda &= 
    \begin{bmatrix}
        1 & 1 & 0 & 0 \\
        0 & 0 & 1 & 1 \\
    \end{bmatrix}.
\end{align*}
Based on the above transportation network model, the user equilibrium model (denoted as problem \textbf{P1}) in optimization form can be formulated as follows:
\begin{align}
    \textbf{P1: } \min\:~&F^\text{ue}= 
    \begin{aligned}[t]
        &\sum_a^{|\mathcal{A}|}\int_0^{x^\text{arc}_a}\omega t_a(\theta)\text{d}\theta \\
        &+\sum_m^{|\mathcal{N}^\text{fcs}|} \left( \int_0^{x^\text{fcs}_{m}}\omega t^\text{fcs}_{m}(\theta)\text{d}\theta + E\lambda_{m}x^\text{fcs}_{m} \right)
    \end{aligned}\label{ueobj} \\
    \text{s.t.:}~&\boldsymbol{x}^\text{arc}=\Delta^{\text{arc}} \boldsymbol{f}, \label{arcflow}\\
    &\boldsymbol{x}^\text{fcs}=\Delta^{\text{fcs}} \boldsymbol{f}, \label{chargeflow}\\
    &\Lambda\boldsymbol{f}=\boldsymbol{D}, \label{demand_e}\\
    &\boldsymbol{f}\geq 0, \label{non-negative}
\end{align}
where symbol $\boldsymbol{f}\in\mathbb{R}^{|\mathcal{P}|} $ is the vector of the path flow. Symbol $t_a$ is the arc travel time of arc $a$ and  $t^\text{fcs}_{m}$ is the charging time of FCS $m$. Symbol $\lambda_{m}$ denotes the charging price of FCS $m$. Symbol $\omega$ is the time cost coefficient, converting the time to money. Arc-path and charge-path incidence relationships are satisfied by constraints (\ref{arcflow})-(\ref{chargeflow}), where $\boldsymbol{x}^\text{arc}=\left[x^\text{arc}_{a}\right]\in\mathbb{R}^{\left|\mathcal{A}\right|}$ is the arc flow vector and $\boldsymbol{x}^{\text{fcs}}=\left[x^\text{fcs}_{m}\right]\in\mathbb{R}^{\left|\mathcal{N}^{\text{fcs}}\right|}$ is the charging flow vector. Flow conservation is held by constraint (\ref{demand_e}). Constraint (\ref{non-negative}) restricts path flow to be non-negative. 

\subsection{Charging Service Pricing Model}
As mentioned at the beginning of this section,  we only optimize one charging service provider, but the model can be expanded into a competitive one via iteration. Each self-interested provider optimizes charging prices for its FCSs to maximize profit (denoted as problem \textbf{P2}), which can be formulated as follows:
\begin{align}
    \textbf{P2:}\quad \max_{\boldsymbol{\lambda}}~&F^\text{csp}=E\boldsymbol{\lambda}^\intercal\Tilde{\boldsymbol{x}}^\text{fcs*},\label{objcsp}\\
    \text{s.t.:}~&\underline{\lambda}\leq\boldsymbol{\lambda}\leq\overline{\lambda}\label{pricebound}.
\end{align}
In the objective function (\ref{objcsp}), $\boldsymbol{\lambda}\in\mathbb{R}^{|\mathcal{N}^\text{fcs}_1|}$ is the charging price vector of FCSs held by the optimized charging service provider, where $\mathcal{N}^\text{fcs}_1\subset\mathcal{N}^\text{fcs}$ is the subset of all FCS set associating to this provider. Symbol $\Tilde{\boldsymbol{x}}^\text{fcs*}\in\mathbb{R}^{|\mathcal{N}^\text{fcs}_1|}$ denotes the charging flow of the optimized FCSs. Constraint (\ref{pricebound}) sets the upper and lower bound for charging prices. It should be noted that in the formulation, we ignore the electricity cost from the power network for brevity, {however, it could be easily added back into the formulation as $(\boldsymbol{\lambda}-\boldsymbol{\nu^*})$, where $\boldsymbol{\nu^*}$ is the locational marginal price vector from the power network. {The detailed operation model of the power network is presented in Appendix \ref{appendix:power network}.}}

In problem \textbf{P2}, $\Tilde{\boldsymbol{x}}^\text{fcs*}$ is a constant vector from the results of the user equilibrium model \textbf{P1}. As a result, the charging service provider cannot capture the response of the charging flow to the charging price variation and, therefore, cannot make the optimal pricing strategy. To this end, it is common in the existing studies to convert problems \textbf{P1} and \textbf{P2} into a single-level problem (denoted as problem \textbf{P3}) via the KKT conditions, which can be written as follows:
\begin{align}
    \textbf{P3:}\quad \max~&F^\text{csp}=E\boldsymbol{\lambda}^\intercal\Tilde{\boldsymbol{x}}^\text{fcs},\label{reOBJCSP}\\
    \text{s.t.:}~&(\ref{demand_e}), (\ref{non-negative}), (\ref{pricebound}),\nonumber\\    &\Delta^{\text{arc}\intercal}\boldsymbol{c}^\text{arc}+\Delta^{\text{fcs}\intercal}\boldsymbol{c}^\text{fcs}-\boldsymbol{\pi}-\Lambda^\intercal\boldsymbol{\mu}=0,\label{kkt1}\\    
    &\boldsymbol{c}^\text{arc}=\omega \boldsymbol{t},\label{carc}\\
    &\boldsymbol{c}^\text{fcs}=\omega \boldsymbol{t}^\text{fcs}+E\boldsymbol{\lambda},\label{cfcs}\\
    &\boldsymbol{\pi}^\intercal \boldsymbol{f}=0,\label{ncc}\\
    &\boldsymbol{\pi}\geq0\label{nonnegativepi},
\end{align}
{where $\boldsymbol{c}^\text{arc}$ is the travel cost on the road, $\boldsymbol{c}^\text{fcs}$ is the cost in the FCS consisting of time and charging price cost.} Constraints (\ref{demand_e})-(\ref{non-negative}), (\ref{kkt1})-(\ref{nonnegativepi}) are the KKT conditions of the user equilibrium model. Symbol $\boldsymbol{t}=[t_a]\in\mathbb{R}^{|\mathcal{A}|}$ and $\boldsymbol{t}^\text{fcs}=[t^\text{fcs}_{m}]\in\mathbb{R}^{|\mathcal{N}^\text{fcs}|}$ are the vectors of arc travel time and charging time. Symbols $\boldsymbol{\pi}\in\mathbb{R}^{|\mathcal{P}|}$ and $\boldsymbol{\mu}\in\mathbb{R}^{|\mathcal{W}|}$ are the vector of Lagrange multipliers associated to constraint (\ref{non-negative}) and constraint (\ref{demand_e}), respectively.

\subsection{Complexities Analysis}\label{complexity analysis}
After reformulation, problem \textbf{P3} becomes a mathematical program with equilibrium constraints. There are three tricky terms in the reformulated problem \textbf{P3}. The first one is the complementary constraint (\ref{ncc}), which is widely linearized via the big-M method. Nevertheless, the big-M method introduces an auxiliary binary variable for each complementary constraint. Since the scale of constraint (\ref{ncc}) is equal to $|\mathcal{P}|$, the number of introduced binary variables increases significantly in a large-scale transportation network. Another issue arises in the objective function (\ref{reOBJCSP}), where a bilinear term occurs. Common methods to relax this term include discretization and the McCormick envelope. Discretization introduces additional binary variables, the number of which depends on its accuracy requirement. Consequently, computational speed is adversely affected due to the increased number of binary variables. McCormick method constructs a convex envelope surrounding the bilinear function with the cost of larger relaxation errors. Two bilinear relaxation methods are limited by either efficiency or optimality. The third issue involves nonlinear time latency functions (see equations (\ref{arcT}) and (\ref{chargeT})), which can only be handled by commercial solvers after applying piecewise linearization techniques (which also introduce binary variables). In conclusion, the reformulated problem \textbf{P3} is generally transformed into a mixed-integer program, with its complexity being exponential to the number of binary variables. This implies that solving this problem in large-scale networks with mathematical programming will result in poor computational performance. 

\section{Gradient Descent Method based on Generalized Sensitivity Analysis}\label{solution method}
According to the above analysis, it is evident that the conventional demand capture method relies on the KKT conditions of the user equilibrium model, leading to various complex terms. However, changes in EV drivers' routing and charging decisions in response to variations in charging prices can be captured through the gradient information from problem \textbf{P1}, thereby avoiding non-convex reformulation. In this section, we adopt the generalized sensitivity analysis to obtain the gradient of charging flows with respect to charging prices $\nabla_{\boldsymbol{\lambda}}\Tilde{\boldsymbol{x}}^\text{fcs}$ and propose a gradient descent algorithm with the obtained gradient information to solve problem \textbf{P2}, which is named the gradient descent method based on generalized sensitivity analysis (GDGSA). 

Considering problem \textbf{P2}, the update process is written as:
\begin{align}
    \boldsymbol{\lambda}_{iter+1}&=\boldsymbol{\lambda}_{iter}+\alpha_{iter}\boldsymbol{h}_{iter},\label{priceupdate}
\end{align}
where $\alpha_{iter}$ is the optimal stepsize and $\boldsymbol{h}_{iter}$ is the feasible direction vector. If inequality (\ref{maxCondition}) and the price bounding constraint (\ref{pricebound}) are simultaneously satisfied for each iteration, then $F^\text{csp}$ at least will reach a local optimum. 
\begin{equation}
    F^\text{csp}(\Tilde{\boldsymbol{x}}^{\text{fcs}}_{iter+1}, \boldsymbol{\lambda}_{iter+1})> F^\text{csp}(\Tilde{\boldsymbol{x}}^{\text{fcs}}_{iter}, \boldsymbol{\lambda}_{iter}).\label{maxCondition}
\end{equation}
According to the update equation (\ref{priceupdate}), this section is structured as follows. In Section \ref{feasible direction}, we introduce the approach to compute the feasible direction $\boldsymbol{h}$. In Section \ref{sensitivity analysis}, we generalize the conventional sensitivity analysis to be compatible with charging behaviors and introduce the calculation process of the gradient $\nabla_{\boldsymbol{\lambda}} \Tilde{\boldsymbol{x}}^\text{fcs}$ based on generalized sensitivity analysis. In Section \ref{stepsize}, we introduce a method to determine the optimal stepsize $\alpha$. In Section \ref{solution summary}, we give a comprehensive process of the proposed pricing method with an illustrative case to help readers understand, and analyze the theoretical performance of the proposed method, GDGSA.

\subsection{Feasible Direction Method}\label{feasible direction}
The charging service pricing model \textbf{P2} is a constrained optimization problem. To avoid charging prices out of the feasible region, we adopt a norm-relaxed method proposed by Cawood and Kostreva \cite{cawood1994normrelaxed} to determine the feasible direction. We can obtain the feasible search direction by solving the following quadratic program:
\begin{align}
    \textbf{P4:}\quad\max_{z,\boldsymbol{h}}~&z-\frac{\gamma}{2}\boldsymbol{h}^\intercal \boldsymbol{Q}\boldsymbol{h},\label{obj:fdd}\\
    \text{s.t.:}~&-\nabla_{\boldsymbol{\lambda}} F^{\text{csp}\intercal}\boldsymbol{h}+z\leq0,\label{cons:dircetion}\\
    &\boldsymbol{\lambda}-\overline{\lambda}+\boldsymbol{h}+z\leq0,\label{cons:ineq1}\\
    &\underline{\lambda}-\boldsymbol{\lambda}-\boldsymbol{h}+z\leq0,\label{cons:ineq2}
\end{align}
where parameter $\gamma$ is a positive constant, whose value may affect the convergence speed of the gradient descent\cite{cawood1994normrelaxed}. The optimal selection of parameter $\gamma$ is beyond the scope of this paper and omitted here. Symbol $\boldsymbol{Q}$ is any given positive-definite matrix, e.g., the identity matrix. Symbol $z$ is the decision variable (scalar). Note that constraints (\ref{cons:ineq1}) and (\ref{cons:ineq2}) are both compact form, which contains $2\times|\mathcal{N}^\text{fcs}_1|$ constraints. The gradient $\nabla_{\boldsymbol{\lambda}} F^\text{csp}$ of objective function (\ref{reOBJCSP}) in constraint (\ref{cons:dircetion}) is given by the following form:
\begin{equation}
    \nabla_{\boldsymbol{\lambda}} F^\text{csp}=E[(\nabla_{\boldsymbol{\lambda}} \Tilde{\boldsymbol{x}}^\text{fcs})^\intercal \boldsymbol{\lambda}+\Tilde{\boldsymbol{x}}^\text{fcs}].
\end{equation}
The above quadratic program can be efficiently solved by commercial solvers. Note that the feasible direction determination requires the gradient $\nabla_{\boldsymbol{\lambda}} \Tilde{\boldsymbol{x}}^\text{fcs}$ as input information, which is introduced in the next subsection.

\subsection{Generalized Sensitivity Analysis}\label{sensitivity analysis}
We adopt the sensitivity analysis method to obtain $\nabla_{\boldsymbol{\lambda}} \Tilde{\boldsymbol{x}}^\text{fcs}$. The sensitivity analysis was first proposed by Tobin and Friesz in 1988 as a general approach to obtain the response of the arc flow $\boldsymbol{x}^\text{arc}$ to a certain disturbance\cite{tobin1988sensitivity}, which had been widely adopted in optimal toll design\cite{yan1996optimal}, traffic control\cite{yang1994traffic}, etc.  However, the original sensitivity analysis method had been questioned in terms of the validity and generality\cite{josefssonpitfalls}. Specifically, the original sensitivity analysis method is strongly restricted by the topology of the transportation network, which means it fails to work correctly in some cases. To address this, two types of more general sensitivity analysis methods were proposed: reduced-arc-based\cite{cho2000reductiona} and reduced-path-based\cite{yang2007sensitivitya} sensitivity analysis. {Two different sensitivity analyses can generate the same gradient, however the reduced-arc-based sensitivity analysis requires more time consumption on finding the maximal linear independent groups, and more complicated matrix construction. Therefore, we use the reduced-path-based sensitivity analysis in this study.} 

\begin{figure}
    \centering
    \includegraphics[scale=0.5]{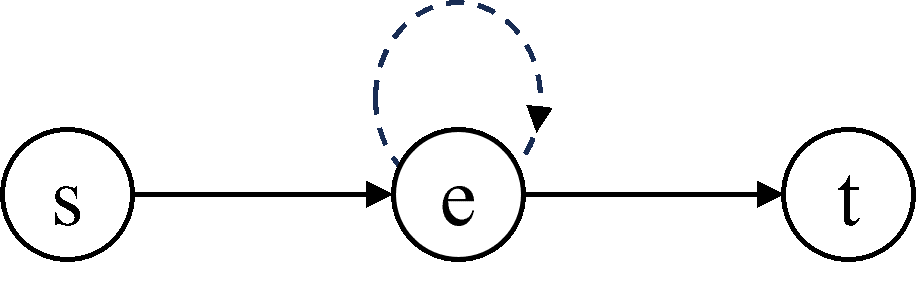}
    \caption{Hyper-arc transformation.}
    \label{fig:hyper arc}
\end{figure}

Another issue is that the conventional sensitivity analysis cannot involve the EVs' charging behaviors. To handle this, we propose a generalized sensitivity analysis method, which converts a charging behavior on the node into a generalized travel behavior on the arc. We demonstrate this conversion in Fig.~\ref{fig:hyper arc}. Symbols $s$ and $t$ are any nodes before and after the charging node $e\in\mathcal{N}^\text{fcs}$, respectively. When an EV driver recharges on the node $e$, it is converted to drive across the dashed arc, which is called the ``hyper arc" since it is not a real road, and the ``travel cost" across the hyper arc is equal to $\boldsymbol{c}^\text{fcs}$, which is shown in equation (\ref{cfcs}). After the transformation, the arc flow $\boldsymbol{x}^\text{arc}$ and the charging flow $\boldsymbol{x}^\text{fcs}$ can be considered as a generalized arc flow $\boldsymbol{x}=[\boldsymbol{x}^\text{arc};\boldsymbol{x}^\text{fcs}]\in\mathbb{R}^{|\mathcal{A}|+|\mathcal{N}^\text{fcs}|}$, and the arc-path $\Delta^\text{arc}$ and charge-path matrix $\Delta^\text{fcs}$ can be similarly written as a generalized arc-path matrix $\Delta=[\Delta^\text{arc};\Delta^\text{fcs}]\in\mathbb{R}^{|\mathcal{A}|+|\mathcal{N}^\text{fcs}|\times|\mathcal{P}|}$. Furthermore, we can write the generalized arc cost as $\boldsymbol{c}^\text{garc}=[\boldsymbol{c}^\text{arc};\boldsymbol{c}^\text{fcs}]$. The advantage of this transformation is that it does not change the mathematical property of the user equilibrium problem, the generalized formulation can involve charging behaviors with mild assumption (i.e. EVs recharge only once) and facilitate the proofs later. 

To improve the readability of the derivation for obtaining $\nabla_{\boldsymbol{\lambda}} \Tilde{\boldsymbol{x}}^\text{fcs}$, we rewrite the KKT conditions of the user equilibrium model as follows:
\begin{align}
     &\boldsymbol{c}-\boldsymbol{\pi}-\Lambda^\intercal\boldsymbol{\mu}=0,\label{kkt11}\\    
    &\boldsymbol{\pi}^\intercal \boldsymbol{f}=0,\label{ncc1}\\
    &\Lambda\boldsymbol{f}=\boldsymbol{D},\label{demandconser}\\
    &\boldsymbol{\pi},\boldsymbol{f}\geq0\label{nonnegativepif},
\end{align}
where $\boldsymbol{c}=\Delta^\intercal\boldsymbol{c}^\text{garc}$, we convert the generalized arc cost to the path cost. In our study, the charging price $\boldsymbol{\lambda}$ is regarded as the disturbance, then we should ensure that $\boldsymbol{x}$ and $\boldsymbol{f}$ are unique and continuous in the neighborhood of the disturbance, so that the implicit function theorem can be correctly adopted\cite{yang2007sensitivitya}. We first prove the generalized arc flows $\boldsymbol{x}$ are unique through \textbf{Proposition~\ref{prop:uniqueness}}.

\begin{proposition}\label{prop:uniqueness}
    Suppose the generalized arc cost is strictly monotone and twice differentiable in $\boldsymbol{x}$, then $\nabla_{\boldsymbol{x}}\boldsymbol{c}^\text{garc}$ is positive-definite and the user equilibrium model has the unique solution $\boldsymbol{x}$.
\end{proposition}

\noindent\textit{Proof.} Since the constraints of the user equilibrium model are convex, the generalized arc flow $\boldsymbol{x}$
is unique if the Hessian matrix of the objective function to $\boldsymbol{x}$ is positive-definite. According to reference \cite{sheffi1984urban}, the Hessian matrix is given by (\ref{delCX}). Based on (\ref{equ:par_carc})-(\ref{equ:par_cfcs}) and the assumption of the \textbf{Proposition~\ref{prop:uniqueness}}, all diagonal entries of (\ref{delCX}) are strictly positive and all non-diagonal entries are zero, thus it is positive-definite and $\boldsymbol{x}$ is unique.\qed
\begin{equation}
\begin{split}
     \nabla_{\boldsymbol{x}}^2F^\text{ue}&=\nabla_{\boldsymbol{x}}\boldsymbol{c}^\text{garc}
     \\&=\begin{bmatrix}
\nabla_{\boldsymbol{x}^\text{arc}}\boldsymbol{c}^\text{arc} & 0 \\
0 & \nabla_{\boldsymbol{x}^\text{fcs}}\boldsymbol{c}^\text{fcs}
\end{bmatrix}, \label{delCX}
\end{split}
\end{equation}
\begin{equation}
   \frac{\partial c^\text{arc}_i}{\partial x^\text{arc}_j}=
   \begin{cases}
       \omega\frac{dt_i}{dx^\text{arc}_i},&~i=j\\
       0,&~i\neq j
   \end{cases},\label{equ:par_carc}
\end{equation}
\begin{equation}
        \frac{\partial c^\text{fcs}_i}{\partial x^\text{fcs}_j}=
   \begin{cases}
       \omega\frac{dt^\text{fcs}_i}{dx^\text{fcs}_i},&~i=j\\
       0.&~i\neq j
   \end{cases}.\label{equ:par_cfcs}
\end{equation}

\noindent\textit{Remark.} The assumption of the \textbf{Proposition~\ref{prop:uniqueness}} can be easily satisfied since it is common to use the BRP function (\ref{arcT}) and the cubic function (\ref{chargeT})\cite{liu2021multiagent} as arc travel and FCSs charging time latency functions, respectively: 
\begin{align}
    t_a&=t_a^0(1+0.15(x^\text{arc}_a/u_a)^4),\forall a\in\mathcal{A},\label{arcT}\\
    t^\text{fcs}_{m}&=t^\text{fcs0}_{m}+\overline{t}^\text{fcs}_{m}(x^\text{fcs}_{m}/u^\text{fcs}_{m})^3, \label{chargeT}\forall m\in\mathcal{N}^\text{fcs},
\end{align}
where $t_a^0$ and $u_a$ are the free travel time and road capacity of arc $a$, respectively. Similarly, $t^\text{fcs}_m$ and $u^\text{fcs}_m$ are the free charging time and the capacity of FCS $m$, respectively. Symbol $\overline{t}^\text{fcs}_m$ is the waiting time coefficient. In our study, we adopt the solution method based on the cone relaxation from Sheng \emph{et al.}\cite{sheng2023impact}, which can fast and accurately solve the user equilibrium problem \textbf{P1}.

Then we simplify the KKT system (\ref{kkt11})-(\ref{nonnegativepif}) for further derivation. The uniqueness of the path flow $\boldsymbol{f}$ will be discussed later. 

In problem \textbf{P1}, path flows are all assigned to the paths with the lowest cost. In other words, there is no path flow on the paths with higher costs. According to this, we define the ``Non-equilibrated path" (NEP) set as $\mathcal{P}^\text{NEP}=\{p|c_p>c_p^*,\forall p\in\mathcal{P}\}$, thus we have $f_p=0, \forall p\in\mathcal{P}^\text{NEP}$. Here symbol $c_p^*$ represents the lowest path cost of the OD pair associated with path $p$. Therefore, the NEP set does not affect the solution of user equilibrium and we can remove these paths. We add ``$\hat{~}$" on all related vectors and matrices to denote that NEPs are removed, and we call the remaining paths as ``Equilibrated path" (EP), and the corresponding set is $\mathcal{P}^\text{EP}=\mathcal{P}- \mathcal{P}^\text{NEP}$. After removing NEPs, it is noted that $\pi_p=0,f_p\geq0,\forall p\in\mathcal{P}^\text{EP}$. Therefore, we can ignore constraints (\ref{ncc1}) and (\ref{nonnegativepif}), then the KKT system is modified as follows:
\begin{align}
    \hat{\boldsymbol{c}}-\hat{\Lambda}^\intercal\boldsymbol{\mu}=0,\label{kktEP}\\
    \hat{\Lambda}\hat{\boldsymbol{f}}=\boldsymbol{D}.\label{demandEP}
\end{align}
Then, we should further restrict the network so that the generalized sensitivity analysis is general for any transportation topology. Before doing so, we first explain in detail why the original sensitivity analysis has limitations, which helps understand the motivation of the latter network restriction process. In the original sensitivity analysis, Tobin and Friesz \cite{tobin1988sensitivity} suppose that $[\hat{\Delta};\hat{\Lambda}]$ is a full column rank matrix. If their opinion is correct, then according to implicit function theorem\cite{tobin1988sensitivity}, we have: 
\begin{align}\begin{bmatrix}\nabla_{\boldsymbol{\lambda}}\hat{\boldsymbol{f}}\\ \nabla_{\boldsymbol{\lambda}}\boldsymbol{\mu}\end{bmatrix}&=-\boldsymbol{J}_{\hat{\boldsymbol{f}},\boldsymbol{\mu}}^{-1}\boldsymbol{J}_{\boldsymbol{\lambda}},\label{implicitfunction}
\end{align}
where $\boldsymbol{J}_{\hat{\boldsymbol{f}},\boldsymbol{\mu}}$ and $\boldsymbol{J}_{\boldsymbol{\lambda}}$ are the Jacobi matrices of the KKT system (\ref{kktEP})-(\ref{demandEP}) with respect to both of EP flow and Lagrange multipliers, and charging price (or say disturbance), respectively. They are given by:
\begin{align}
        \boldsymbol{J}_{\hat{\boldsymbol{f}},\boldsymbol{\mu}}&=\begin{bmatrix}
        \nabla_{\hat{\boldsymbol{f}}}\hat{\boldsymbol{c}}&-\hat{\Lambda}^\intercal\\ \hat{\Lambda}&0
    \end{bmatrix},\label{Jfu}\\
    \boldsymbol{J}_{\boldsymbol{\lambda}}&=\begin{bmatrix}
        \nabla_{\boldsymbol{\lambda}}\hat{\boldsymbol{c}}\\\nabla_{\boldsymbol{\lambda}}\boldsymbol{\mu}
    \end{bmatrix},\label{Jlambda}
\end{align}
since $\hat{\boldsymbol{c}}=\hat{\Delta}^\intercal\boldsymbol{c}^\text{garc}$ and $\boldsymbol{x}=\Delta\boldsymbol{f}$, then $\frac{\partial\hat{\boldsymbol{c}}}{\partial\hat{\boldsymbol{f}}}=\frac{\partial\hat{\Delta}^\intercal\boldsymbol{c}^\text{garc}}{\partial\boldsymbol{x}}\frac{\partial\boldsymbol{x}}{\partial\hat{\boldsymbol{f}}}$. Therefore, $\nabla_{\hat{\boldsymbol{f}}}\hat{\boldsymbol{c}}=\hat{\Delta}^\intercal\nabla_{\boldsymbol{x}}\boldsymbol{c}^\text{garc}\hat{\Delta}$. In (\ref{Jlambda}), since $\nabla_{\boldsymbol{\lambda}}\boldsymbol{c}^\text{arc}=0$ and $\nabla_{\boldsymbol{\lambda}}\hat{\boldsymbol{c}}=\hat{\Delta}^{\text{arc}\intercal}\nabla_{\boldsymbol{\lambda}}\boldsymbol{c}^\text{arc}+\hat{\Delta}^{\text{fcs}\intercal}\nabla_{\boldsymbol{\lambda}}\boldsymbol{c}^\text{fcs}$, then $\nabla_{\boldsymbol{\lambda}}\hat{\boldsymbol{c}}=\hat{\Delta}^{\text{fcs}\intercal}\nabla_{\boldsymbol{\lambda}}\boldsymbol{c}^\text{fcs}$. Similarly, $\boldsymbol{\mu}$ is not the function of the charging price $\boldsymbol{\lambda}$, so $\nabla_{\boldsymbol{\lambda}}\boldsymbol{\mu}=0$. Thus, equation (\ref{Jlambda}) can be simplified as follows:
\begin{equation}
    \boldsymbol{J}_{\boldsymbol{\lambda}}=\begin{bmatrix}        \hat{\Delta}^{\text{fcs}\intercal}\nabla_{\boldsymbol{\lambda}}\boldsymbol{c}^\text{fcs}\\0
    \end{bmatrix}.\label{re_Jlambda}
\end{equation}
We can convert the gradient of EP flow with respect to the charging price to that of charging flow with respect to the charging price by the following equation:
\begin{equation}
    \nabla_{\boldsymbol{\lambda}}\Tilde{\boldsymbol{x}}^\text{fcs}=\hat{\Delta}^\text{fcs}\nabla_{\boldsymbol{\lambda}}\hat{\boldsymbol{f}}.
\end{equation}
Since there exists matrix inversion in the implicit function theorem equation (\ref{implicitfunction}), it raises concern about the invertibility of the Jacobi matrix. The invertibility can be guaranteed through \textbf{Proposition~\ref{inverse}}.
\begin{proposition}\label{inverse}
    Suppose $[\hat{\Delta};\hat{\Lambda}]$ is a full column rank matrix, then $\boldsymbol{J}_{\hat{\boldsymbol{f}},\boldsymbol{\mu}}$ is invertible\cite{yang2007sensitivitya}.  
\end{proposition}

\noindent\textit{Proof.} The matrix is invertible if its null space only contains the zero vector. Therefore, we construct the homogeneous linear equations $\boldsymbol{J}_{\hat{\boldsymbol{f}},\boldsymbol{\mu}}\boldsymbol{\eta}=0$, where $\boldsymbol{\eta}=[\boldsymbol{\eta}_1;\boldsymbol{\eta}_2]\in\mathbb{R}^{|\mathcal{P}^\text{EP}|+|\mathcal{W}|}$. If $\boldsymbol{\eta}$ is unique and equal to zero, $\boldsymbol{J}_{\hat{\boldsymbol{f}},\boldsymbol{\mu}}$ is invertible. We expand the linear equations as follows:
\begin{align}
    &\nabla_{\hat{\boldsymbol{f}}}\hat{\boldsymbol{c}}\boldsymbol{\eta}_1-\hat{\Lambda}^\intercal\boldsymbol{\eta}_2=0,\label{inverseProof1}\\
    &\hat{\Lambda}\boldsymbol{\eta}_1=0\label{inverseProof2}.
\end{align}
Since $\nabla_{\hat{\boldsymbol{f}}}\hat{\boldsymbol{c}}=\hat{\Delta}^\intercal\nabla_{\boldsymbol{x}}\boldsymbol{c}^\text{garc}\hat{\Delta}$, we substitute this equation into equation (\ref{inverseProof1}) and left time $\boldsymbol{\eta}_1^\intercal$:
\begin{equation}
    (\hat{\Delta}\boldsymbol{\eta}_1)^\intercal\nabla_{\boldsymbol{x}^\text{garc}}\boldsymbol{c}\hat{\Delta}\boldsymbol{\eta}_1-(\hat{\Lambda}\boldsymbol{\eta}_1)^\intercal\boldsymbol{\eta}_2=0.
\end{equation}
Note that $\hat{\Lambda}\boldsymbol{\eta}_1=0$ and $\nabla_{\boldsymbol{x}}\boldsymbol{c}^\text{garc}$ is positive-definite as given in \textbf{Proposition~\ref{prop:uniqueness}}. According to the property of positive-definite matrix, $\hat{\Delta}\boldsymbol{\eta}_1=0$. Then we have:
\begin{equation}
    \begin{bmatrix}\hat{\Delta}\\ \hat{\Lambda}    \end{bmatrix}\boldsymbol{\eta}_1=0.
\end{equation}
Since $[\hat{\Delta};\hat{\Lambda}]$ is a full column rank matrix, then $\boldsymbol{\eta}_1$ is unique and $\boldsymbol{\eta}_1=0$. In equation (\ref{inverseProof1}),  $\nabla_{\hat{\boldsymbol{f}}}\hat{\boldsymbol{c}}\boldsymbol{\eta}_1=0$, now we have $\hat{\Lambda}^\intercal\boldsymbol{\eta}_2=0$. According to the definition of OD-path matrix (see Section.~\ref{uemodel}), $\hat{\Lambda}^\intercal$ must be full column rank, thus $\boldsymbol{\eta}_2=0$ as well. Then we yield that null space of $\boldsymbol{J}_{\hat{\boldsymbol{f}},\boldsymbol{\mu}}$ only contains zero vector, so it is invertible.\qed

Now we see that the original sensitivity analysis is based on the incorrect condition: $[\hat{\Delta};\hat{\Lambda}]$ is a full column rank matrix. The property of full column rank cannot be guaranteed solely through removing the NEP set, thus the invertibility fails to be guaranteed. Therefore, we should further restrict the transportation network.

The basic idea of network restriction is that when $[\hat{\Delta};\hat{\Lambda}]$ is not full column rank, we find the maximal linearly independent columns of the matrix. After that, we can use the maximal linearly independent columns to span the space or say to represent other redundant columns. We define the maximal linearly independent columns as ``Equilibrated linearly independent" (ELI) set, denoted as $\mathcal{P}^\text{ELI}\subseteq\mathcal{P}^\text{EP}$. The linearly dependent paths are denoted as ``ELD" set, which is denoted as $\mathcal{P}^\text{ELD}\subset\mathcal{P}^\text{EP}$. We use subscripts ``1" and ``2" to distinguish ELI and ELD respectively. Now we write the relationship between EP flow and generalized arc flow as follows:
\begin{equation}
    \begin{bmatrix}\hat{\Delta}_1&\hat{\Delta}_2\\ \hat{\Lambda}_1&\hat{\Lambda}_2 \end{bmatrix}\begin{bmatrix} \hat{\boldsymbol{f}}_1\\ \hat{\boldsymbol{f}}_2   \end{bmatrix}=\begin{bmatrix}\boldsymbol{x}\\ \boldsymbol{D}    \end{bmatrix}.\label{pathflowwitharcflow}
\end{equation}
It is noted that if the left-hand side coefficient matrix in (\ref{pathflowwitharcflow}) can be decomposed into the above form, then finding the ELI set is necessary. Otherwise, it means that $[\hat{\Delta};\hat{\Lambda}]$ is full column rank, then the whole problem degenerates into the situation discussed in the original sensitivity analysis. Now the uniqueness of path flow is guaranteed as well. Moreover, we can just use the ELI set in the process of sensitivity analysis since the ELI set has contained all the required information, and the ELD set can be represented by the linear combination of the ELI set. Then the KKT system now is given by:
\begin{align}
    \hat{\boldsymbol{c}}_1-\hat{\Lambda}_1^\intercal\boldsymbol{\mu}=0,\\
    \hat{\Lambda}_1\hat{\boldsymbol{f}}_1=\boldsymbol{D},
\end{align}
where $\hat{\boldsymbol{c}}_1=\hat{\Delta}_1^\intercal\nabla_{\boldsymbol{x}}\boldsymbol{c}^\text{garc}$. The remaining derivation is the same as the above process since $[\hat{\Delta}_1;\hat{\Lambda}_1]$ is guaranteed to be a full column rank matrix, and $\boldsymbol{J}_{\hat{\boldsymbol{f}}_1,\boldsymbol{\mu}}$ is invertible.
\subsection{Optimal Stepsize Determination}\label{stepsize}
In optimal stepsize determination of the adopted feasible direction method\cite{cawood1994normrelaxed}, it solves a linear program with stepsize as the decision variable. However, the optimal stepsize determination by solving a linear program is not available for our problem since the charging flow is an implicit function of the charging price. {Common strategies include fixed or adaptive step sizes. Fixed step sizes lack flexibility: a large step may accelerate calculation but risk slow convergence, while a small step wastes time. Adaptive step sizes require tuning the decay rate, which is challenging due to varying objective function landscapes across transportation topologies, necessitating problem-specific adjustments.} Therefore, we adopt a method based on ``trial-error," the basic idea of which is to gradually increase the stepsize until inequality (\ref{maxCondition}) and price bounding constraint (\ref{pricebound}) fail to hold. The steps are shown as in \textbf{Algorithm~\ref{arg1}} (suppose we have obtained the feasible direction $\boldsymbol{h}$). Although \textbf{Algorithm~\ref{arg1}} solves problem \textbf{P1} for up to $\overline{k}$ times in one iteration, the overall efficiency is still acceptable. On the one hand, problem \textbf{P1} is a convex problem, which can be efficiently solved by off-the-shelf solvers. On the other hand, though there is no theoretical proof, from our experience, the $k$ will be close to $\overline{k}$ only for the first few iterations. As a result, the GDGSA is still computationally efficient.
\begin{algorithm}
\caption{Stepsize Determination}
\label{arg1}
\begin{algorithmic}[1]
    \STATE Set the basic stepsize $\alpha^0$ and $\overline{k}$, and let $k=1$
    \REPEAT
    \STATE Let $\alpha=k\alpha^0$, and update charging price through equation (\ref{priceupdate})
    \STATE Solve the \textbf{P1} with updated charging price
    \IF{condition (\ref{maxCondition}) and constraint (\ref{pricebound}) are satisfied}
    
    \STATE k=k+1
    \STATE Continue
    \ELSE
    \STATE Output $\alpha$ as optimal stepsize
    \STATE Break
    \ENDIF
    \UNTIL  $k>\overline{k}$
\end{algorithmic}\label{alg:stepsize}
\end{algorithm}

\subsection{Solution Algorithm Summary}\label{solution summary}
\subsubsection{Overall Solution Process}
The whole solution process of the optimal pricing problem is summarized in \textbf{Algorithm~\ref{alg:grad descent}}.
\begin{algorithm}
\caption{Gradient Descent Algorithm}
\label{arg2}
\begin{algorithmic}[1]
    \STATE Initialization: input $\Delta^\text{arc},\Delta^\text{fcs},\Lambda,\boldsymbol{D}$, all parameters, set initial charging prices, $iter=1$ and convergence tolerance $\epsilon$
    \REPEAT
    \STATE Solve problem \textbf{P1} to obtain $\boldsymbol{x}^\text{arc}$ and $\boldsymbol{x}^\text{fcs}$
    \STATE Calculate $\nabla_{\boldsymbol{\lambda}}\Tilde{\boldsymbol{x}}^\text{fcs}$ based on generalized sensitivity analysis
    \STATE Solve problem \textbf{P4} to find the feasible direction $\boldsymbol{h}_{iter}$
    \STATE Find the optimal stepsize $\alpha_{iter}$ via \textbf{Algorithm~\ref{alg:stepsize}}
    \STATE Update charging prices with equation (\ref{priceupdate})
    \IF{$(F^{\text{csp}}_{iter}-F^{\text{csp}}_{iter-1})\leq\epsilon$}
    
    \STATE Break;
    \ELSE
    \STATE $iter=iter+1$
    \ENDIF
    \UNTIL  $iter>\overline{iter}$
\end{algorithmic}\label{alg:grad descent}
\end{algorithm}

\subsubsection{An Illustrative Case}
We use a small case shown in Fig.~\ref{illustration} to help readers better understand how the generalized sensitivity analysis works. The input matrices and traffic demand have been given before. Assume the generalized arc cost functions are all $1+x$. The optimized FCS is located on the node $II$ (Another located at $III$ is its rival) and the charging prices of FCS located on $II\text{ and }III$ are both 1 at first. Then we directly give the results of the user equilibrium problem: $\boldsymbol{f}=[0.75,0.75,1,1]^\intercal$, $\boldsymbol{x}^\text{arc}=[1.75,0.75,1.75,1,1,0.75]^\intercal$, $\boldsymbol{x}^\text{fcs}=[1.75,1.75]^\intercal$, $\boldsymbol{c}=[8.25,8.25,8.5,8.5]^\intercal$, $\boldsymbol{c}^*=[8.25,8.25,8.5,8.5]^\intercal$. Firstly, we find that the $\mathcal{P}^\text{EP}\equiv\mathcal{P}$ since all paths are equilibrated ($\boldsymbol{c}\equiv\boldsymbol{c}^*$). Secondly, we check the property of $[\hat{\Delta};\hat{\Lambda}]$ and its rank is 4, thus it is a full column rank matrix and $\hat{\Delta}_2,\hat{\Lambda}_2=\emptyset$. Then the sensitivity analysis degenerates to the original one (i.e. finding maximal linear independent group is not needed). Thirdly, the Jacobi matrix $\boldsymbol{J}_{\hat{\boldsymbol{f}},\boldsymbol{\mu}}$ is given by:
$$
\boldsymbol{J}_{\hat{\boldsymbol{f}},\boldsymbol{\mu}}=\begin{bmatrix}
    \begin{array}{cccc:cc}
        3&0 &2&0&-1&0 \\
        0& 3&0&2&-1&0\\
        2& 0&3&0&0& -1\\
        0&2 &0&3&0&-1\\ \hdashline
        1&1 &0&0&0& 0\\
        0& 0&1&1&0&0\\
    \end{array}
\end{bmatrix}.
$$
The matrix is divided into four blocks corresponding to the four blocks in equation (\ref{Jfu}). Then the $\boldsymbol{J}_{\boldsymbol{\lambda}}$ is given by:
$$
\boldsymbol{J}_{\boldsymbol{\lambda}}=
\begin{bmatrix}
    \begin{array}{c}
        1\\
        0\\
        1\\
        0\\ \hdashline
        0\\
        0\\
    \end{array}
\end{bmatrix}.
$$
It should be noted that only FCS located at node $II$ is optimized, thus $\nabla_{\boldsymbol{\lambda}}\boldsymbol{c}^\text{fcs}=[1;0]$ (the charging price of the rival FCS is a constant). Then the gradient can be calculated as follows:
$$
\nabla_{\boldsymbol{\lambda}}\Tilde{\boldsymbol{x}}^\text{fcs}=\hat{\Delta}^\text{fcs}\nabla_{\boldsymbol{\lambda}}\hat{\boldsymbol{f}}=-\hat{\Delta}^\text{fcs}\boldsymbol{J}_{\hat{\boldsymbol{f}},\boldsymbol{\mu}}^{-1}\boldsymbol{J}_{\boldsymbol{\lambda}}=\begin{bmatrix}
    -0.2\\0.2
\end{bmatrix}.
$$
The gradient shows that $x^\text{fcs}_1$ and $x^\text{fcs}_2$ will decrease and increase with the same degree if the $\lambda_1$ increases, respectively. It is straightforward to see that the result is correct.
\subsubsection{Theoretical Analysis of the Algorithm}
From the perspective of computational efficiency, several steps are time-consuming, including solving the user equilibrium model \textbf{P1}, finding the feasible direction \textbf{P4}, and identifying the maximal linearly independent groups. Theoretically, problems \textbf{P1} and \textbf{P4} are both cone programs, which can be solved efficiently by the interior point method in polynomial time. Based on experimental experience, solving problems \textbf{P1} and \textbf{P4} are both very fast, typically within 0.1s and 0.05s, respectively. Finding the inverse of the Jacobi matrix $\boldsymbol{J}_{\hat{\boldsymbol{f}}_1,\boldsymbol{\mu}}$ adopts the LU decomposition in MATLAB with polynomial time complexity. Moreover, there are two factors resulting in faster inversion. On the one hand, $\boldsymbol{J}_{\hat{\boldsymbol{f}}_1,\boldsymbol{\mu}}$ is sparse. On the other hand, there exist two times of network restriction (i.e. removing NEP set and ELD set from the original path set), so most of the ``useless" paths are eliminated, and the size of $\boldsymbol{J}_{\hat{\boldsymbol{f}}_1,\boldsymbol{\mu}}$ will be limited. MATLAB implements the Gauss-Jordan elimination method to find the maximal linear independent groups, which is polynomial time complexity as well. However, the size of the processed matrix $\boldsymbol{J}_{\hat{\boldsymbol{f}},\boldsymbol{\mu}}$ while finding the maximal linear independent groups is far larger than $\boldsymbol{J}_{\hat{\boldsymbol{f}}_1,\boldsymbol{\mu}}$ while calculating inverse. As a result, finding maximal linear independent groups accounts for the main part of time consumption. {In conclusion, the time complexity of the GDGSA is theoretically polynomial depending on the scale of Jacobi matrix $\boldsymbol{J}_{\hat{\boldsymbol{f}},\boldsymbol{\mu}}\in\mathbb{R}^{(|\mathcal{P}^\text{EP}|+|\mathcal{W}|)\times(|\mathcal{P}^\text{EP}|+|\mathcal{W}|)}$, that is the sizes of the path set and OD pairs.}

{From the perspective of optimality, the GDGSA fails to guarantee a global optimal solution. Due to the intrinsic non-convex property of objective function (\ref{objcsp}) (see a more specific example in Fig.~\ref{fig:csp profit}), the GDGSA is likely to fall into a local optimum and it is affected by the initial point. However, the mathematical programming fails to find the global optimum as well due to the relaxation errors.}

{From the perspective of convergence, although the problem \textbf{P3} is non-convex, we can prove that the proposed algorithm converges to the local optima, see details in Appendix \ref{appendix:convergence}.}

In summary, the proposed algorithm has significant efficiency enhancement and acceptable optimal value loss compared to mathematical programming.

\section{Experiments}\label{result}
In this section, we select three transportation networks of varying sizes (small, medium, and large) to verify the algorithmic performance of GDGSA, Nguyen-Dupuis \cite{cui2023multiperiod}, Eastern Massachusetts \cite{zeng2023conic}, and a modified Winnipeg network based on \cite{stabler2024bstabler} (see network configuration in Table~\ref{tab:Network Configuration}). {The Eastern Massachusetts is an inter-city highway network, which we couple with an 135 kV IEEE-30 node power network. The Winnipeg is an intra-city road network, for which the 138 kV is applied. The model parameters are set as follows: charging energy $E$ 50 kWh, time cost $\omega$ 2 \$/hour, lower and upper bounds of charging price 200 and 230 \$.} The parameters of the GDGSA in the three cases are the same, where $\gamma$=2, $\boldsymbol{Q}$ is set as an identity matrix, algorithm tolerance $\epsilon$=1e-3, basic stepsize $\alpha^0$=1, $\overline{k}$=50. In the algorithm, there are several steps involving solving mathematical programs, where the user equilibrium model is solved by the MOSEK solver\cite{mosek} and the feasible direction problem is solved by the GUROBI solver\cite{gurobi}. All coding works are based on MATLAB R2023b (the mathematical programmings are modeled by YALMIP toolbox\cite{lofberg2004yalmip}) on a laptop with i5-13500HX CPU.

To show the computational efficiency, we use the mathematical programming (denoted as MP in the subsequent Tables and analysis) to solve problem \textbf{P3} as the benchmark, where the bilinear objective function (\ref{objcsp}) is relaxed by the McCormick envelope, the complementary constraints (\ref{ncc}) are addressed via the big-M method, and the nonlinear time latency functions (\ref{arcT})-(\ref{chargeT}) are piecewise linearized. {The relaxed mixed integer problem is solved by the GUROBI solver. The Gurobi parameters are set as follows: \textit{MIPGap} is set to 1e-3, \textit{TimeLimit} is set to 259,200 (3 days), \textit{MIPFocus} is set to 1 to accelerate solution. The \textit{MIPFocus} can force the solver to find feasible integer first, which can effectively improve the efficiency of the model with large-scale integers.}
 \begin{table}[]
\caption{Network Configuration}
\centering
\begin{tabular}{ccccc}
\hline
Network  & OD & Node &Arc&FCS\\ \hline
Nguyen-Dupuis & 4  & 13&18&4 \\
Eastern Massachusetts& 1113&74&258&41  \\
Winnipeg & 1373 &1057&2535&97    \\ \hline
\end{tabular}\label{tab:Network Configuration}
\end{table}

\subsection{Nguyen-Dupuis}
\begin{figure}
    \centering
    \includegraphics[scale=0.5]{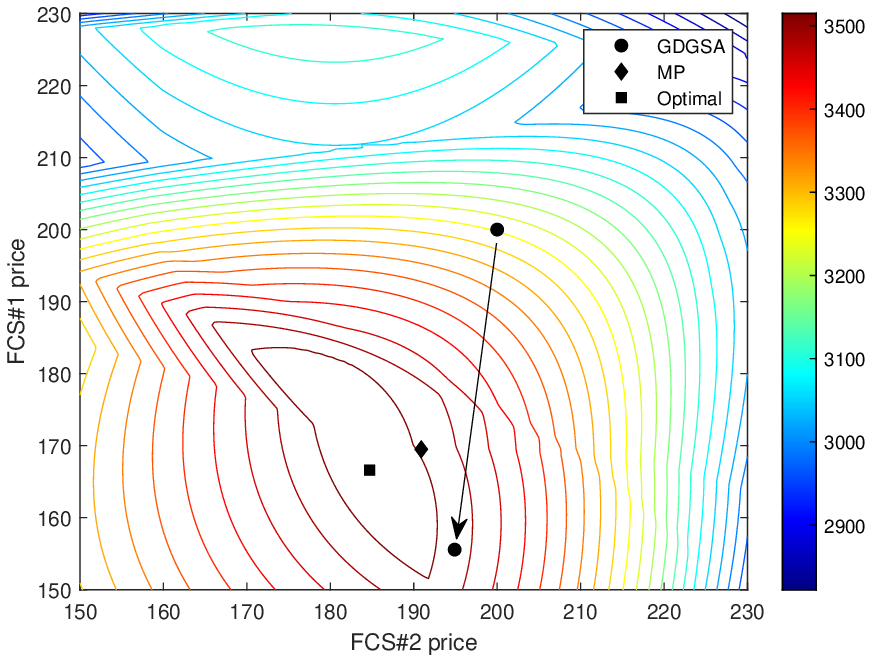}
    \caption{Charging service provider's profit function visualization.}
    \label{fig:csp profit}
\end{figure}

In order to intuitively observe the mathematical properties of the pricing problem and the performance of the solving methods, we enumerate possible charging price strategies and solve problem \textbf{P1} for 25600 times with different charging prices to plot Fig.~\ref{fig:csp profit}. In Fig.~\ref{fig:csp profit}, the two circles denote the initial and optimal points found by the GDGSA, which are connected by an arrow, the diamond is the solution of the MP and the square is the real optimal point. It is obvious that the profit function is non-convex and there are several ridges, thus the GDGSA is easy to drop to a local maximum. {The reason why the MP cannot find the global optimum is that the piecewise linearization technique and McCormick envelope are implemented, thus there exist relaxation errors. In Table~\ref{tab:performance nguyen}, the profit gaps of the two methods compared to the optimal profit are 0.3\% and 1.4\%, respectively, while the GDGSA costs 38.3\% less time than the MP.  }
\begin{table}[]
\caption{Algorithm Performance in Nguyen-Dupuis}
\centering
\begin{tabular}{ccc}
\hline
Method  & Profit (\$) & Solution time (s) \\ \hline
MP      & 1062   & 11.5 \\
GDGSA      & 1050   & 7.1  \\
Optimal & 1065   & -    \\ \hline
\end{tabular}\label{tab:performance nguyen}
\end{table}
\subsection{Eastern Massachusetts}
Different from the Nguyen-Dupuis, we cannot enumerate all available paths in such a scale network, thus we {generate finite different scales of path sets} to demonstrate the performance (see detailed path generation algorithm in reference \cite{pan2024competitive}). {We first evaluate the algorithm's performance by comparison to the mathematical programming method, where several path sets with different scales are generated to test the scalability of the proposed algorithm. We then evaluate the impact of the optimal price strategy with other fixed strategies on the charging service provider's profit and the coupled power and transportation networks.}

In Table~\ref{tab:performance massachusetts}, the GDGSA shows a significant advantage in terms of computational efficiency in the four path sets, where it saves 67.3\%, 83.9\%, and 91.4\% computation time for the first three path set sizes compared to the MP, and the MP fails to complete the solution in {259,200} seconds for the last one. It is noted that the increment of time consumption of mathematical programming is rapid with the path set scale expanding, while that of the proposed algorithm shows mild growth. As we analyzed in Section~\ref{solution summary}, the complexity of mathematical programming is exponential to the number of binary variables. Lots of binary variables are introduced to relax the complementary constraints (\ref{ncc}), resulting in additional time consumption. However, the proposed GDGSA first restricts the network by removing all NEP sets and further restricts the network by finding the maximal linear independent groups, thus it will not be affected by those useless paths and still maintain high efficiency. For optimal profits, the GDGSA shows +0.8\%, +1.5\%, and -4.5\% profit gaps compared to mathematical programming, respectively. Although the GDGSA can find a better profit in the first two path sets, it cannot show the robust ability to find a better charging price strategy all the time. However, tuning the basic stepsize $\alpha^0$ may help find a better profit with the consequence of longer calculation time, thus it is a trade-off between profit and efficiency.
\begin{table}[]
\caption{Algorithm Performance in Eastern Massachusetts}
\centering
\begin{tabular}{cccccc}
\hline
Method               & Path set size & 2775   & 3182   & 4039   & 6085   \\ \hline
\multirow{2}{*}{MP} & Profit (\$)  & 1410.2 & 1443.1 & 1543.5 & -     \\
                     & Solution time (s) & 216.4  & 641.6  & 1570.1 & -     \\ \hline
\multirow{2}{*}{GDGSA}  & Profit (\$)  & 1421.3 & 1464.1 & 1473.6 & 1473.9 \\
                     & Solution time (s) & 70.8   & 102.9  & 134.5  & 144.6  \\ \hline
\end{tabular}\label{tab:performance massachusetts}
\end{table}

From the perspective of the social operator, they may raise concerns about whether the optimal pricing of the charging service will deteriorate the operation of the coupled network. Therefore, we carry out the numerical experiment to evaluate the impact of the optimal pricing on the coupled network. Since we cannot obtain the real power grid data of eastern Massachusetts, we adopt the IEEE-30 network instead as the power network for theoretical analysis purposes. We first solve the charging service provider pricing problem and convert the charging flow $\boldsymbol{x}^\text{fcs}$ into power load via function $E\boldsymbol{x}^\text{fcs}/\boldsymbol{t}^\text{fcs0}$, then solve the optimal power flow through MATPOWER\cite{zimmerman2011matpower}. The locational marginal price from the optimal power flow is sent back to the charging service provider as electricity cost and repeats until converges. {The detailed operation model of the power network and the interaction process are provided in Appendix \ref{appendix:power network}.} We set three fixed charging price strategies as benchmarks, the prices of which are set to lower bound, mean, and upper bound (i.e. let $\boldsymbol{\lambda}=\underline{\lambda}\text{, }(\underline{\lambda}+\overline{\lambda})/2\text{, }\overline{\lambda} $, respectively). The interaction results are presented in Table~\ref{tab:impact in massachusetts}, the profit under the optimal pricing is better than the fixed price strategies without a doubt. The impact on the coupled network is not evident in this case. For the overall transportation network time, which has been converted into monetary cost by coefficient $\omega$, the Optimal strategy spends 0.6\% more time cost than the Mean strategy, while the power network loss is the lowest.
\begin{table}[]
\caption{Impact of Pricing Strategy on the Coupled Network (Eastern Massachusetts)}
\centering
\begin{tabular}{cccc}
\hline
Strategy    & Profit & PN loss (MW) & TN cost (\$) \\ \hline
Optimal     & 459.3  & 2.71         & 3665.2       \\
Lower bound & 419.9  & 2.73         & 3660.2       \\
Mean        & 290.1  & 2.91         & 3642.5       \\
Upper bound & 196.0  & 2.86         & 3644.9       \\ \hline
\end{tabular}\label{tab:impact in massachusetts}
\end{table}
\subsection{Winnipeg}
{Similarly, we first evaluate the algorithm's performance in terms of optimality, efficiency, and scalability in several path sets with different scales. We then evaluate the impact of the optimal pricing with other fixed strategies on the charging service provider's profit and the coupled networks. }Table~\ref{tab:performance winnipeg} summarizes the profits and solution times with four path sets with different scales of the Winnipeg network by mathematical programming and the proposed GDGSA. {Mathematical programming fails to solve cases of the last two path sets in 259,200 seconds, while the GDGSA can still handle them within a maximum of 500 seconds. Although the GDGSA shows 9.3\% and 10.0\% profit gap compared to the MP, it shows far faster computational speed, up to maximal 222 times, than the MP. The high computation efficiency indicates the potential of the GDGSA in the urban-scale calculation.}

For the impact analysis in the coupled network, the power network is the IEEE-118 network and the interaction framework is the same as that in the above case. The results are presented in Table~\ref{tab:impact in winnipeg}. The profit comparison is still significant, the charging service provider under the Optimal strategy can earn 62.3 times more profit than the Upper bound strategy. {Different from eastern Massachusetts, the Optimal strategy obviously shows the negative effect on the coupled network, where it leads to 5.3\% more power network loss and 2.6\% more transportation network time than the Mean strategy.} For the charging service provider, the best profit is the trade-off between quantity (charging demand) and price. For some FCSs, the Optimal strategy asks them to lower the price and attract more customers, thus the power load of the related power buses increases. As a result, the uneven power load leads to more power loss. However, this ``small profits but more quick turnover" phenomenon will not occur in all optimized FCSs, which is also the reason why the power loss is not increased in the eastern Massachusetts case. For the transportation network, the additional time cost is due to the detouring to the cheaper FCSs for recharging and waiting time in the crowded FCSs. In this case, we can conclude that the charging service provider is a self-interested entity and it is likely to undermine the operation of the coupled network. 
\begin{table}[]
\caption{Algorithm Performance in Winnipeg}
\centering
\begin{tabular}{cccccc}
\hline
Method               & Path set size & 3996   & 5678   & 7497   & 9200   \\ \hline
\multirow{2}{*}{MP} & Profit (\$)  & 2140.2 & 2486.2 & - & -     \\
                     & Solution time (s) & 15632.8  & 83412.6  & - & -     \\ \hline
\multirow{2}{*}{GDGSA}  & Profit (\$)  & 1941.7 & 2238.2 & 2270.6 & 2374.3 \\
                     & Solution time (s) & 172.6   & 375.8  & 432.4  & 493.5  \\ \hline
\end{tabular}\label{tab:performance winnipeg}
\end{table}
\begin{table}[]
\caption{Impact of Pricing Strategy on the Coupled Network (Winnipeg)}
\centering
\begin{tabular}{cccc}
\hline
Strategy    & Profit & PN loss (MW) & TN cost (\$) \\ \hline
Optimal     & 1894.6  & 79.6         & 9734.8       \\
Lower bound & 1760.9  & 78.4         & 9624.5       \\
Mean        & 679.2  & 76.1         & 9487.3       \\
Upper bound & 30.4  & 75.6         & 9501.4       \\ \hline
\end{tabular}\label{tab:impact in winnipeg}
\end{table}
\section{Conclusion}\label{conclusion}
{This paper introduces the sensitivity analysis technique to the charging service pricing problem and generalizes it to consider EVs' charging behaviors. The detailed proof and calculation process of the generalized sensitivity analysis are presented. Based on the gradient from the generalized sensitivity analysis, a comprehensive iterative solution framework GDGSA is proposed, including the user equilibrium solution method, the generalized sensitivity analysis technique, the feasible direction method, and the optimal stepsize determination.}

{The effectiveness of the proposed algorithm and its impact on power and transportation networks are comprehensively validated through multiple networks, including the Nguyen-Dupuis, Eastern Massachusetts, and Winnipeg. Benchmark results demonstrate that the proposed algorithm achieves superior computational efficiency, outperforming mathematical programming by up to 222 times in computational speed while maintaining a modest optimality gap of approximately 10\%. The optimal pricing strategy demonstrates remarkable economic benefits, generating up to 62.3 times greater profits compared to fixed pricing schemes. However, this enhanced profitability comes with marginal trade-offs, including a 5.3\% increase in power network losses and a 2.6\% rise in transportation network operational costs.}

{In the future work, the algorithm generalization for the dynamic pricing scenario will be taken into consideration. Besides, the uncertainties in the model will be explored. Furthermore, the vehicle-to-grid service of large-scale EVs can be considered in the pricing problem.}





\appendices 
\section{} \label{appendix:power network}
{The power network operation solves the operation model considering the charging power load, outputting the locational marginal price to the charging service providers. Then we yield the optimal pricing strategy as well as the charging demand $\boldsymbol{x}^\text{fcs}$ by solving the optimal pricing problem \textbf{P3}. It should be noted that the objective function of the pricing problem \textbf{P3} is changed into $E(\boldsymbol{\lambda}-\boldsymbol{\nu}^*)^\intercal \Tilde{\boldsymbol{x}}^\text{fcs}$, where the symbol $\boldsymbol{\nu}^*$ is the locational marginal price, the superscript star denotes it is not variable. Afterward, the charging demand can be transformed into the power load via the equation $\boldsymbol{p}^\text{fcs}=E\boldsymbol{x}^\text{fcs}/\boldsymbol{t}^\text{fcs0}$. The iteration between the power network and the charging service provider is summarized in Algorithm \ref{alg:iteration}.}
\begin{algorithm}
\caption{{Iteration Between Power Network and Charging Service Provider}}
\label{arg3}
\begin{algorithmic}[1]
    \REPEAT
    \STATE {Solve problem \textbf{P5} to obtain locational marginal price}
    \STATE {Solve problem \textbf{P3} to obtain charging demand $\boldsymbol{x}^\text{fcs}$}
    \STATE {Transform charging demand into power load through function $E\boldsymbol{x}^\text{fcs}/\boldsymbol{t}^\text{fcs0}$}

    \UNTIL  {$|\Delta F^\text{csp}|\leq\epsilon$}
\end{algorithmic}\label{alg:iteration}
\end{algorithm}

{The operation model of the power network is denoted as \textbf{P5}, and formulated as follows:}
\begin{align}
    \textbf{P5: }&\min\: \sum_{j\in B^{\text{G}}}(c_{2,j}p_{j}^{\text{G }2}+c_{1,j}p^\text{G }_{j})+c_0p^\text{G }_{0}, \label{objPDN}\\
    \text{s.t.:}~\:&p^\text{L}_{ij}+p^\text{G}_{j}=\sum_{k\in \mathcal{B}^j}p^\text{L}_{jk}+p_{j}^{\text{d}}+p_{j}^{\text{fcs}}, \forall(i,j)\in \mathcal{L}, \label{pbalance}\\
    &q^\text{L}_{ij}+q^\text{G}_{j}=\sum_{k\in \mathcal{B}^j}q^\text{L}_{jk}+q_{j}^{\text{d}}, \forall(i,j)\in \mathcal{L}, \label{qbaalance}\\
    &U_{j}=U_{i}-2(r_{ij}p^\text{L}_{ij}+x_{ij}q^\text{L}_{ij})+(z_{ij})^{2}I_{ij}, \forall(i,j)\in \mathcal{L}, \label{voltagedrop}\\
    &(p^\text{L}_{ij})^{2}+(q^\text{L}_{ij})^{2}\leq U_{i}I_{ij},~\forall(i,j)\in \mathcal{L}, \label{SOCP}\\
    &p_{j}^{\text{fcs}}=Ex^\text{fcs}_j/t^\text{fcs0}_j,~\forall j\in \mathcal{B}^{\text{fcs}},\label{coupleCons}\\
    &\underline{U_{j}}\leq U_{j} \leq \overline{U_{j}}, \forall j\in \mathcal{B},\label{Urange}\\    
    &\underline{I_{ij}}\leq I_{ij} \leq \overline{ I_{ij}}, \forall (i,j)\in \mathcal{L}.\label{Irange}
\end{align}

{The power network is represented by the bus set $\mathcal{B}$ and line set $\mathcal{L}$, where symbols $\mathcal{B}^\text{G}$ and $\mathcal{B}^\text{fcs}$ are the subset of the bus set, indicating buses with generators and fast charging stations, respectively. The objective function minimizes the overall generation and injected power from the main grid. Constraints (\ref{pbalance})-(\ref{qbaalance}) denote the node active and reactive power balance. Constraint (\ref{voltagedrop}) depicts the voltage drop across the line. Constraint (\ref{SOCP}) is the second-order cone relaxation of the power function. Constraint (\ref{coupleCons}) reveals the relationship between charging power and charging demand. Constraints (\ref{Urange})-(\ref{Irange}) sets the lower and upper bounds for voltage and current. Since the problem \textbf{P5} is a second-order cone program, the dual variable can be easily obtained from the solver, wherein the dual variable of the active power balance constraint is the locational marginal price $\boldsymbol{\nu}$.}

\section{} \label{appendix:mixed}
{We directly start from the simplified KKT system, where all NEPs are removed. In the following KKT system, we add additional superscripts ``ev" and ``gv" to denote EV and gasoline vehicle.}
\begin{align}
    \hat{\boldsymbol{c}}^\text{ev}-\hat{\boldsymbol{\Lambda}}^{\text{ev},\intercal}\boldsymbol{\mu}^\text{ev}=0,\\
    \hat{\boldsymbol{c}}^\text{gv}-\hat{\boldsymbol{\Lambda}}^{\text{gv},\intercal}\boldsymbol{\mu}^\text{gv}=0,\\    \hat{\boldsymbol{\Lambda}}^{\text{ev}}\hat{\boldsymbol{f}}^\text{ev}=\boldsymbol{D}^\text{ev},\\    \hat{\boldsymbol{\Lambda}}^{\text{gv}}\hat{\boldsymbol{f}}^\text{gv}=\boldsymbol{D}^\text{gv}.
\end{align}

{With the above KKT system, we can follow the similar steps in the main text to calculate the derivatives. It should be noted that only the gradient of EV traffic flow is the required value.}
\begin{align}
    \begin{bmatrix}\nabla_{\boldsymbol{\lambda}}\hat{\boldsymbol{f}}^\text{ev}\\ 
    \nabla_{\boldsymbol{\lambda}}\hat{\boldsymbol{f}}^\text{gv}\\
    \nabla_{\boldsymbol{\lambda}}\boldsymbol{\mu}^\text{ev}\\
    \nabla_{\boldsymbol{\lambda}}\boldsymbol{\mu}^\text{gv}
    \end{bmatrix}&=-\boldsymbol{J}_{\hat{\boldsymbol{f}}^\text{ev},\hat{\boldsymbol{f}}^\text{gv},\boldsymbol{\mu}^\text{ev},\boldsymbol{\mu}^\text{gv}}^{-1}\boldsymbol{J}_{\boldsymbol{\lambda}},
\end{align}
{where two Jacobi matrices are formulated as follows. Substitute the following equations into the above equation, then we can yield the required gradient in the model with mixed EVs and gasoline vehicles.}
\begin{align}
    \boldsymbol{J}_{\hat{\boldsymbol{f}}^\text{ev},\hat{\boldsymbol{f}}^\text{gv},\boldsymbol{\mu}^\text{ev},\boldsymbol{\mu}^\text{gv}}&=
    \begin{bmatrix}
        \nabla_{\hat{\boldsymbol{f}}^\text{ev}}\hat{\boldsymbol{c}}^\text{ev}&\nabla_{\hat{\boldsymbol{f}}^\text{gv}}\hat{\boldsymbol{c}}^\text{ev}&-\hat{\Lambda}^{\text{ev},\intercal}&0\\         \nabla_{\hat{\boldsymbol{f}}^\text{ev}}\hat{\boldsymbol{c}}^\text{gv}&\nabla_{\hat{\boldsymbol{f}}^\text{gv}}\hat{\boldsymbol{c}}^\text{gv}&0&-\hat{\Lambda}^{\text{gv},\intercal}\\
        \hat{\Lambda}^\text{ev}&0&0&0\\
        0&\hat{\Lambda}^\text{gv}&0&0
    \end{bmatrix},\\
    \boldsymbol{J}_{\boldsymbol{\lambda}}&=\begin{bmatrix}
        \nabla_{\boldsymbol{\lambda}}\hat{\boldsymbol{c}}^\text{ev}\\
        \nabla_{\boldsymbol{\lambda}}\hat{\boldsymbol{c}}^\text{gv}\\
        0\\
        0
    \end{bmatrix}.
\end{align}

\section{} \label{appendix:convergence}
{We write the main problem \textbf{P2} here in the minimizing form as follows:}
\begin{align}
    \min_{\boldsymbol{\lambda\in\mathcal{X}}}-F^\text{csp}=-E\boldsymbol{\lambda}^\intercal \Tilde{\boldsymbol{x}}^\text{fcs}(\boldsymbol{\lambda}),\label{obj:appendix proof}
\end{align}
{where the $\mathcal{X}$ is the feasible region of $\lambda$. Here, we change the mathematical form of charging demand into $\Tilde{\boldsymbol{x}}^\text{fcs}(\boldsymbol{\lambda})$ to denote it is a function related to charging prices.}

{We then introduce the concept of Lipschitz continuity. Given a problem $f(x)$, it is Lipschitz continuous if $||f(y)-f(x)||\leq L(y-x),\exists L\geq0$, where $L$ is Lipschitz constant. In our problem, $\nabla_{\boldsymbol{\lambda}} F^\text{csp}=E[(\nabla_{\boldsymbol{\lambda}} \Tilde{\boldsymbol{x}}^\text{fcs})^\intercal \boldsymbol{\lambda}+\Tilde{\boldsymbol{x}}^\text{fcs}]$, the charging demand continuously changes with the charging price, thus the objective function is Lipschitz continuous. Considering the practical factors, the objective function is obviously bounded. For a function with Lipschitz continuity, we have following inequality:}
\begin{align}
    -\nabla^2F^\text{csp}\leq LI.\label{cons:appendix lipschitz}
\end{align}
{The above inequality means the second-order gradient of the objective function is bounded. We then expand the function (\ref{obj:appendix proof}) into second-order form and transform it with the inequality (\ref{cons:appendix lipschitz}):}
\begin{equation}
\begin{split}
    -F^\text{csp}(y) &= -F^\text{csp}(x) - \nabla_{\boldsymbol{\lambda}}F^\text{csp}(x)^\intercal (y-x) \\
    &\quad - \frac{1}{2}(y-x)^\intercal \nabla^2_{\boldsymbol{\lambda}}F^\text{csp}(x)(y-x),\\
     &\leq -F^\text{csp}(x) - \nabla_{\boldsymbol{\lambda}}F^\text{csp}(x)^\intercal (y-x)\\
     &\quad + \frac{L}{2}||y-x||^2,
\end{split}
\end{equation}
{where the symbols $x$ and $y$ are placeholder here to denote the any point within $\mathcal{X}$. }

{Substituting the updating function $\boldsymbol{\lambda}_{iter+1}=\boldsymbol{\lambda}_{iter}+\alpha_{iter}\boldsymbol{h}_{iter}$ into above inequality, we yield:}
\begin{align}
    F^\text{csp}(\boldsymbol{\lambda_{iter+1}})\geq F^\text{csp}(\boldsymbol{\lambda_{iter}})+\alpha_{iter}(1-\frac{L\alpha_{iter}}{2})||\boldsymbol{h}_{iter}||^2_2.\label{descent proof}
\end{align}

{According to the inequality (\ref{descent proof}), if $0\leq\alpha_{iter}\leq\frac{2}{L}$, the algorithm will find higher profit over iteration until gradient $\boldsymbol{h}_{iter}$ or (and) update stepsize $\alpha_{iter}$ approaching to 0. Additionally, owing to the stepsize determination method introduced in Section \ref{stepsize}, we can heuristically ensure the stepsize will not exceed $\frac{2}{L}$. Because though we cannot know the value of $L$, the ``trial-error" stepsize determination can ensure inequality (\ref{maxCondition}) holds, thus the stepsize is within the available range. However, due to the non-convex property of the original problem \textbf{P3}, the algorithm can only converge to the local optima.} 

 

\bibliographystyle{ieeetr}
\bibliography{SA}

\newpage

 




\vfill

\end{document}